\documentclass[12pt]{iopart}
\usepackage{graphicx}
\begin{document}

\null
\noindent

\vskip -2.9cm

\title{From QCD color coherence to inclusive distributions and correlations in jets}

\author{Redamy P\'erez-Ramos}

\address{Dr. Moliner 50, S-46100 Burjassot, Spain}
\ead{redamy.perez@uv.es}

\begin{abstract}
In this paper we briefly review some examples of inclusive energy-momentum distributions and
correlations in QCD intra-jet cascades. Emphasis is given to the role of gluon coherence 
effects in final states hadron spectra. These observables provide tests of the 
Local Parton Hadron Duality (LPHD) hypothesis.
\end{abstract}

High energy annihilation $e^+e^-$ into hadrons, deep ineslatic lepton-hadron
scattering (DIS) and hadron-hadron collisions are classical examples of hard 
processes where high transverse momentum jets are produced \cite{Dokshitzer:1991wu}.
In fact, the observation of quark and gluon jets has played a crucial role
in establishing Quantum Chromodynamics (QCD) as the theory of strong interaction within
the Standard Model of particle physics. Jets, narrowly collimated bundles of hadrons produced
at high virtuality $Q$, reflect configurations of quarks and gluons at short distances.
As is well known, perturbative QCD (pQCD) controls the relevant observables to be measured at colliders
but its applicability fails as the evolution of the jet reaches the hadronization stage. 
In other words, at short quark-gluon distances, the perturbative 
approach is suitable due to the weak strength of the coupling constant 
$\alpha_s(Q^2)\ll1$ while, as the jet evolves towards
hadronization occurring at stronger $\alpha_s(k_\perp^2)\sim1$ 
($k_\perp^2<Q^2$ for secondary partons emitted off the leading parton) and larger quark-gluon distance, 
the expanded series do not converge any longer. Therefore, the perturbative approach fails
to describing the forthcoming evolution of partons 
into final hadrons that hit the detectors. That is why, after the emission of gluons inside the
jet reaches the infrared cut-off $Q_0$, one advocates the Local Parton Hadron Duality 
hypothesis, which mainly consists in comparing the shape and normalization of the obtained distribution
with the corresponding data sets \cite{Azimov:1984np}. 

Jet physics is mainly dominated by soft gluon bremsstrahlung. As a consequence of QCD color coherence,
the emission of successive soft gluons inside the jet has been demonstrated to form a cascade where 
the emission angles decrease towards the hadronization stage, the so-called
Angular Ordering (AO) \cite{Dokshitzer:1991wu}. Perturbative schemes, like the Double Logarithmic Approximation (DLA)
and the Modified Leading Logarithmic Approximation (MLLA), which allow for the 
resummation of soft-collinear and hard-collinear gluons, have been implemented. 
One of the most impressive predictions of perturbative QCD (pQCD) 
is the existence of the hump-backed plateau (HBP) of the inclusive energy distribution of hadrons, 
later confirmed by experiments at colliders like the LEP \cite{Akrawy:1990ha} and the Tevatron \cite{:2008ec}. 
Within the same formalism, the transverse momentum distribution, or $k_\perp$-spectra of
hadrons produced in $p\bar p$ collisions at center of mass energy $\sqrt{s}=1.96$ TeV 
at the Tevatron \cite{Aaltonen:2008yn}, was well described by MLLA and next-to-MLLA (NMLLA)
predictions inside the validity ranges provided by such schemes, both supported by 
the LPHD \cite{PerezRamos:2005nh,Arleo:2007wn}.
Thus, inclusive observables like the inclusive energy distribution and the 
inclusive transverse momentum $k_\perp$ spectra of hadrons have shown that the perturbative stage of the process 
is dominant and the LPHD hypothesis is successful while treating one-particle inclusive observables.
The study of particle correlations in intrajet cascades, which are less inclusive observables, 
provides a refined test of the partonic dynamics and the LPHD. Two-particle correlations were 
measured by the OPAL collaboration in the $e^+e^-$ annihilation at the $Z^0$ peak, 
that is for $\sqrt{s}=91.2$ GeV at LEP \cite{Acton:1992gd}. 
Though the agreement with theoretical predictions 
turned out to be rather good for the description
of the data, a discrepancy still subsists pointing out a possible failure of 
the LPHD for less inclusive observables or the existence of non-trivial hadronization effects
not regarded in the evolution equations. However, these measurements were redone by the CDF collaboration in $p\bar p$ collisions
at the Tevatron for mixed samples of quark and gluon jets \cite{:2008ec}. The agreement with theoretical predictions 
turned out to be rather good, in particular 
for very soft particles ($x\ll0.1$) with very close energy fractions. 
However, a discrepancy between the OPAL and CDF analysis showed up and still stays unclear.

\begin{figure}[h]
\begin{center}
\includegraphics[height=5.4cm,width=7.5cm]{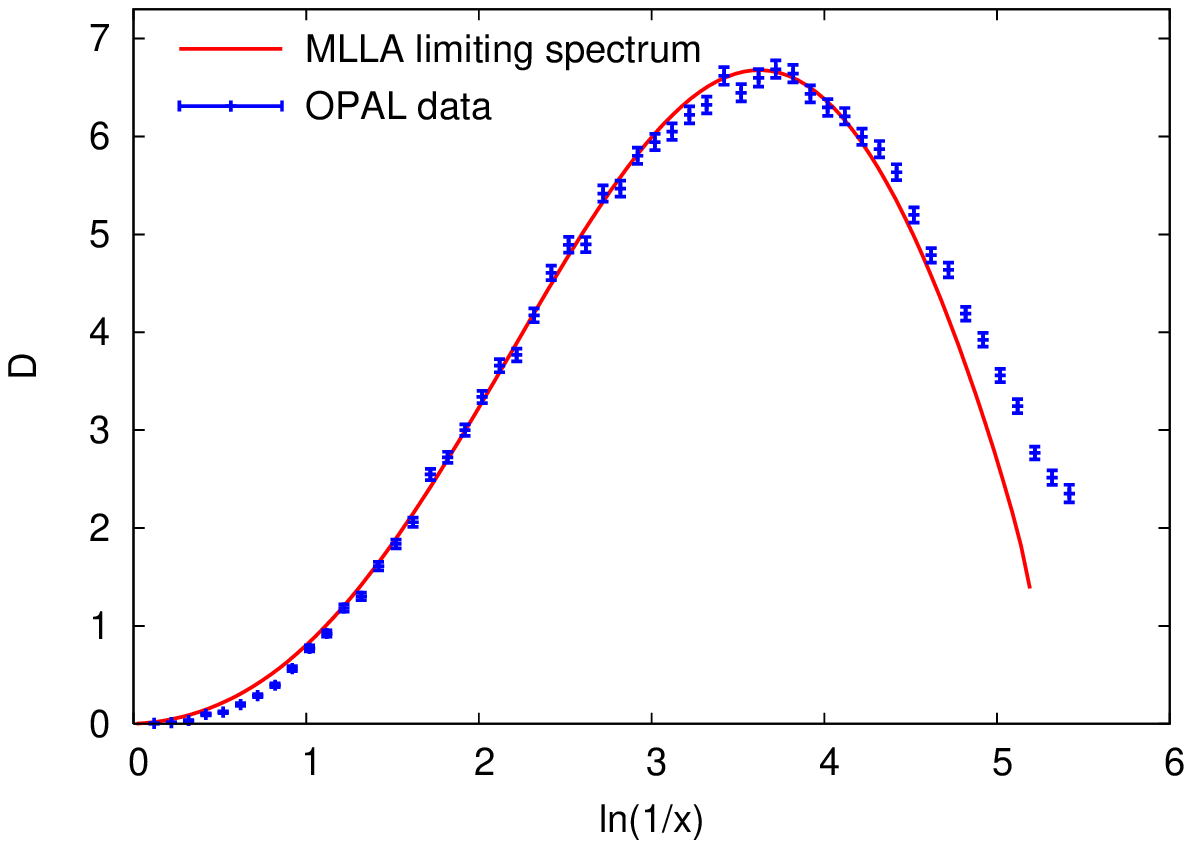}
\includegraphics[height=6.0cm,width=7.5cm]{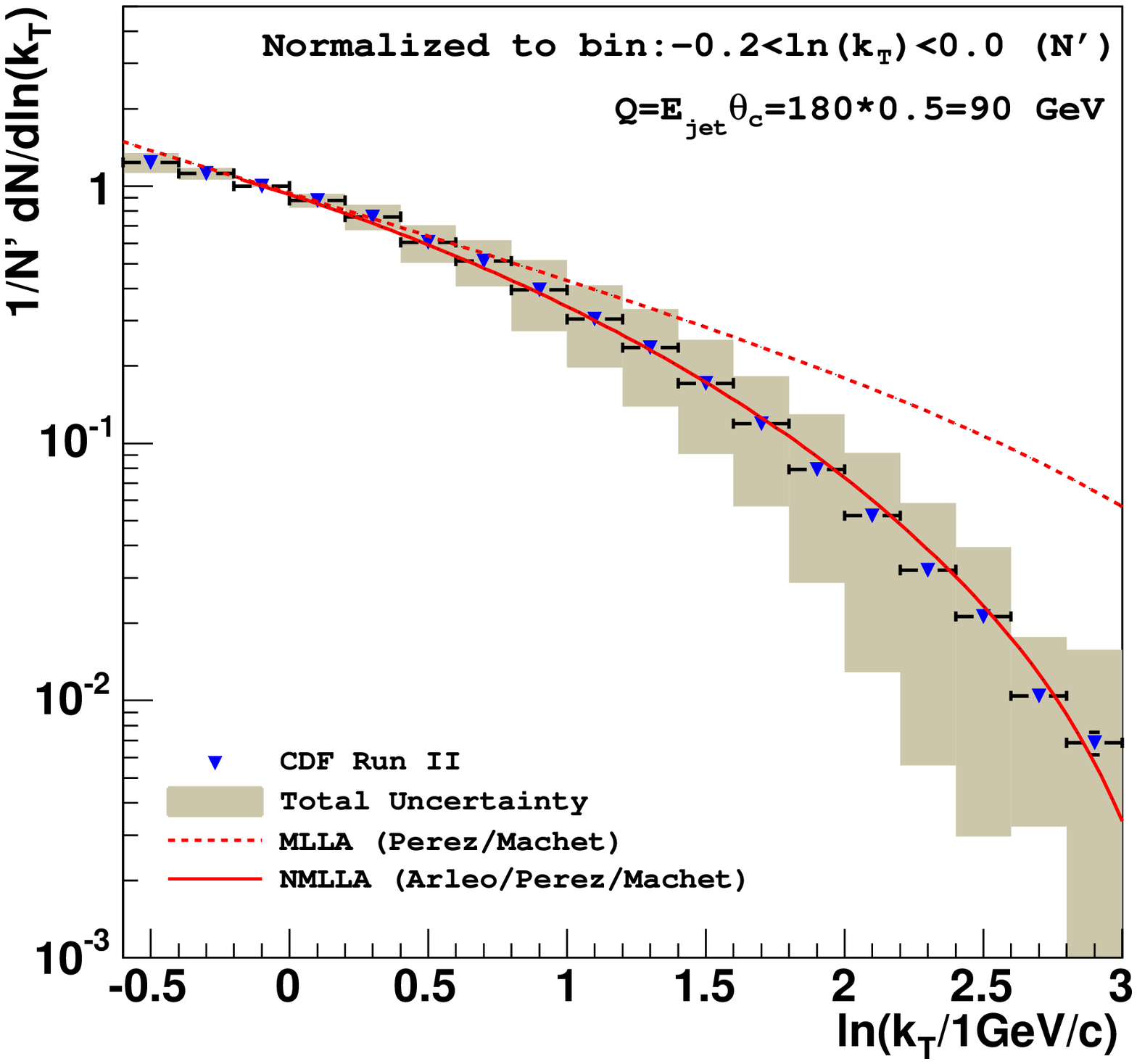}
\caption{\label{fig:one-part-dist} Hump-backed plateau (left) as a function
of $\ln(1/x)$ for fixed $Q=91.2$ GeV in the process $e^+e^-\to q\bar{q}$ 
and $k_\perp$-spectra (right) for a mixed sample of quark and gluon jets
for a dijet with $Q=90$ GeV in $p\bar{p}$ collisions with ${\cal K}^{ch}\sim1$.}  
\end{center}
\end{figure}

The inclusive energy spectrum of soft gluon bremsstrahlung partons in QCD jets 
$D^h(\ln(1/x))={\cal K}^{ch}\frac1{N}\frac{dN}{d\ln(1/x)}$ has been obtained
in the MLLA, and supported with the LPHD hypothesis \cite{Dokshitzer:1991wu}.  
This approximation takes into account all essential ingredients of parton multiplication 
in the next-to-leading order, which are parton splitting functions
responsible for recoil effects at each $q\to qg,\,g\to q\bar q,\,g\to gg$ vertex, the running coupling 
$\alpha_s(k_\perp^2)\ll1$ depending on the relative transverse momentum of the two offspring and the exact AO. 
Gluon coherence suppresses indeed multiple production of very soft gluons such that, only particles with intermediate energies multiply
most efficiently. As a consequence, the energy spectrum of charged hadrons acquires the hump-backed 
shape with an asymptotic energy peak in the logarithmic scale $\ell_{max}=\ln(1/x_{max})\to\frac12\ln(Q/Q_0)$
in the limit where the perturbative approach, regularized by the infrared cut-off $Q_0$, 
is taken down to $\Lambda_{QCD}$, the so-called limiting spectrum. At the end of the day, the shape and normalization
are compared with the experiment; a constant ${\cal K}^{ch}$ is chosen, which normalizes the number of soft gluons to the
number of charged detected hadrons (mostly pions and kaons) turns out to be close to unity (${\cal K}^{ch}\sim1$), 
giving support to the similarity between parton and hadron spectra \cite{Dokshitzer:1991wu}. 
This is exactly what is observed in Fig.\ref{fig:one-part-dist} (left) for the $e^+e^-$ data \cite{Akrawy:1990ha}
at the $Z^0$ peak. 
The description of the OPAL data, to cite one example,
is well described by the MLLA one-particle inclusive distribution in the limiting spectrum approximation. Though the
coupling constant $\alpha_s$ diverges in this limit, the hadron spectra turns out to be Collinear and Infrared Safe (CIS),
describing the data even in the region of $x$, $x\sim1$ where the MLLA is not valid. Other experiments like ALEPH, CDF
DELPHI, HERA and TASSO have also been successful in describing this observable \cite{:2008ec}.

Secondly, the $k_\perp$-spectra of charged hadrons ${\cal K}^{ch}\frac1{N}\frac{dN}{d\ln k_\perp}$ was 
measured at the Tevatron \cite{Aaltonen:2008yn}. It was also proved to follow the 
MLLA and NMLLA expectations and to give further support to the LPHD hypothesis \cite{Arleo:2007wn}.
Computing the single inclusive $k_\perp$-distribution requires
the definition of the jet axis. The starting 
point of the approach consists in considering
the correlation between two particles
(h1) and (h2) of energies $E_1$ and $E_2$ which form a
relative angle $\Theta$ inside one jet of total 
opening angle $\Theta_0>\Theta$ \cite{PerezRamos:2005nh}. 
Weighting over the energy $E_2$ of particle (h2), 
this relation leads to the correlation
between the particle (h=h1)
and the energy flux, which can be identified
with the jet axis \cite{PerezRamos:2005nh}.
Since soft particles are less sensitive to the energy balance, in the 
soft approximation $x\ll1$ the previous correlation disappears and the computation of this observable
becomes straightforward. 
As an example, in Fig.~\ref{fig:one-part-dist} (right), as taken from \cite{Aaltonen:2008yn}, 
we display the $k_\perp$-spectra of charged
hadrons inside a jet of virtuality $Q=90$ GeV in $p\bar{p}$ collisions at
$\sqrt{s}=1.96$~TeV~\cite{Aaltonen:2008yn}, together with the MLLA predictions of
\cite{PerezRamos:2005nh} and the NMLLA calculations \cite{Arleo:2007wn}, both
in the limiting spectrum approximation ($Q_0=\Lambda_{QCD}=230$~MeV);
the experimental distributions suffering from large normalization errors,
data and theory are normalized to the same bin, $\ln(k_\perp/1)=-0.1$.
The results in the limiting spectrum approximation are found to be in
impressive agreement with measurements by the CDF collaboration, unlike
what occurs at MLLA, pointing out small overall
non-perturbative contributions. Moreover, NMLLA predictions are reliable in a much larger
$k_\perp$ range than MLLA. Coherence also plays the same role in this case
at very small $k_\perp$. However, the size of higher order corrections ${\cal O}(\sqrt{\alpha_s})$ as 
$k_\perp\to\Lambda_{QCD}$ becomes huge in such a way that coherence suffers screening due to the running
of $\alpha_s$ \cite{PerezRamos:2005nh}. The agreement between NMLLA predictions
and CDF preliminary data in $p\bar{p}$ collisions at the Tevatron is
good, indicating very small overall non-perturbative corrections and giving further support
to LPHD hypothesis \cite{Azimov:1984np}. 

\begin{figure}[h]
\begin{center}
\includegraphics[height=5.5cm,width=7.5cm]{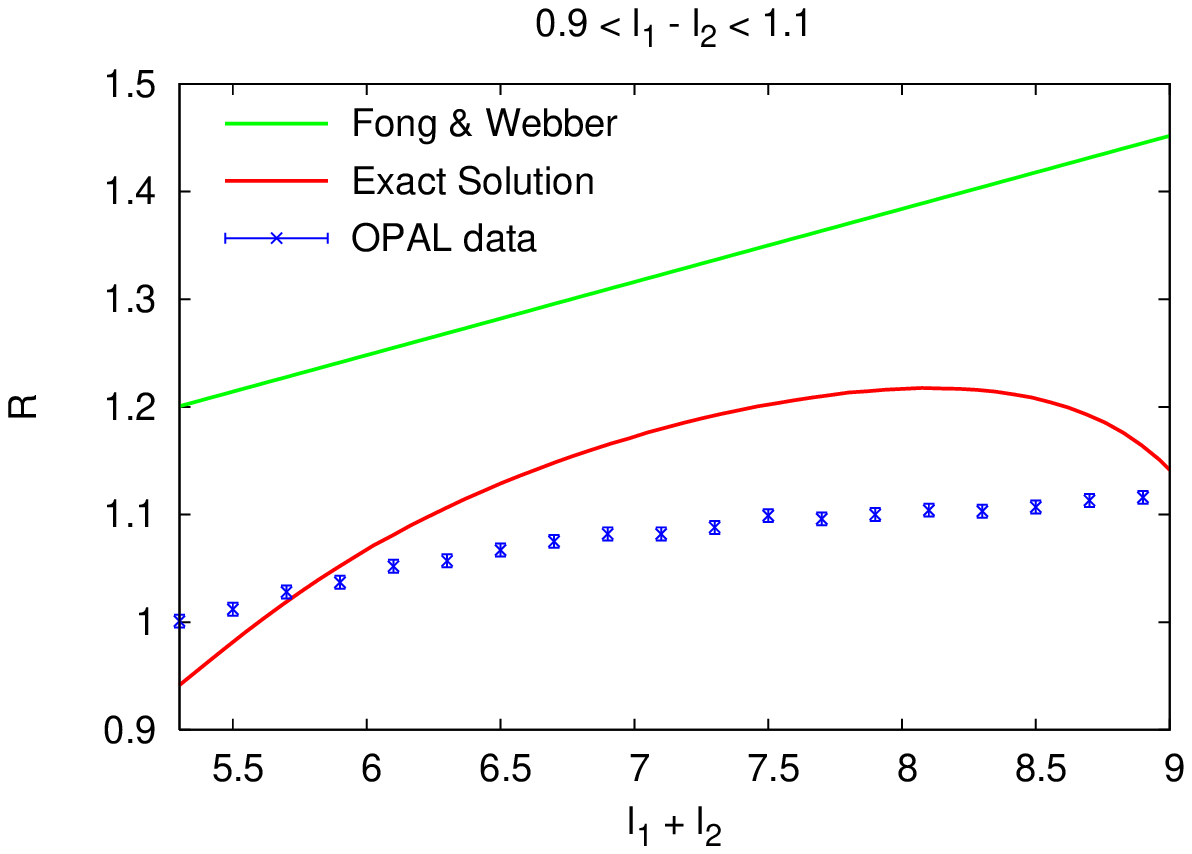}
\includegraphics[height=5.5cm,width=7.5cm]{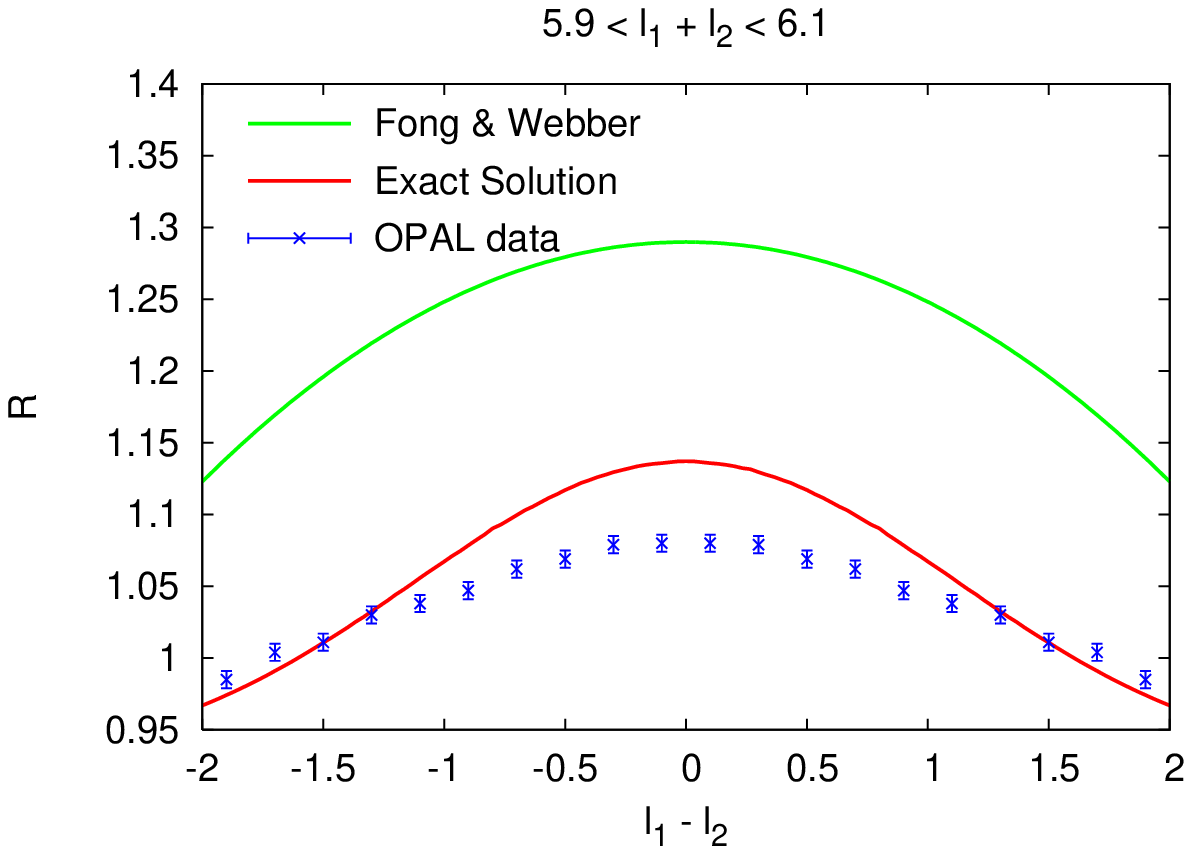}
\caption{\label{fig:opalcorr} Two-particle correlations in two quark 
jets $R=\frac12+\frac12{\cal C}^{(2)}_q$ in the process $e^+e^-\to q\bar{q}$ as a function of 
$\ell_1+\ell_2=|\ln(x_1x_2)|$ for $\ell_1-\ell_2=\ln(x_2/x_1)=1.0$ (left)
and $\ell_1-\ell_2=|\ln(x_2/x_1)|$ for $\ell_1+\ell_2=\ln(x_1x_2)=6.0$ (right).}  
\end{center}
\end{figure}

\begin{figure}[h]
\begin{center}
\includegraphics[height=5.5cm,width=7.5cm]{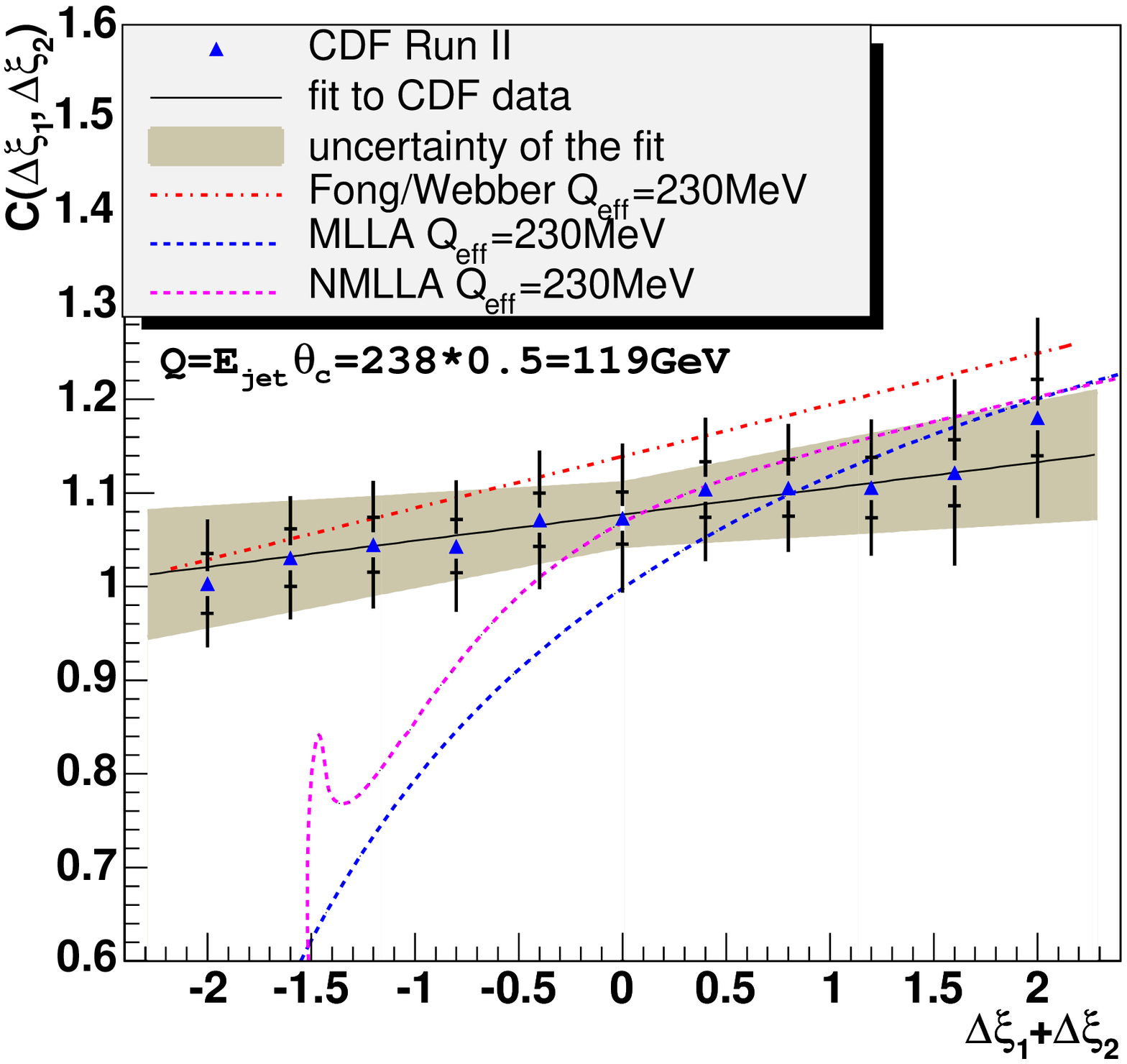}
\includegraphics[height=5.5cm,width=7.5cm]{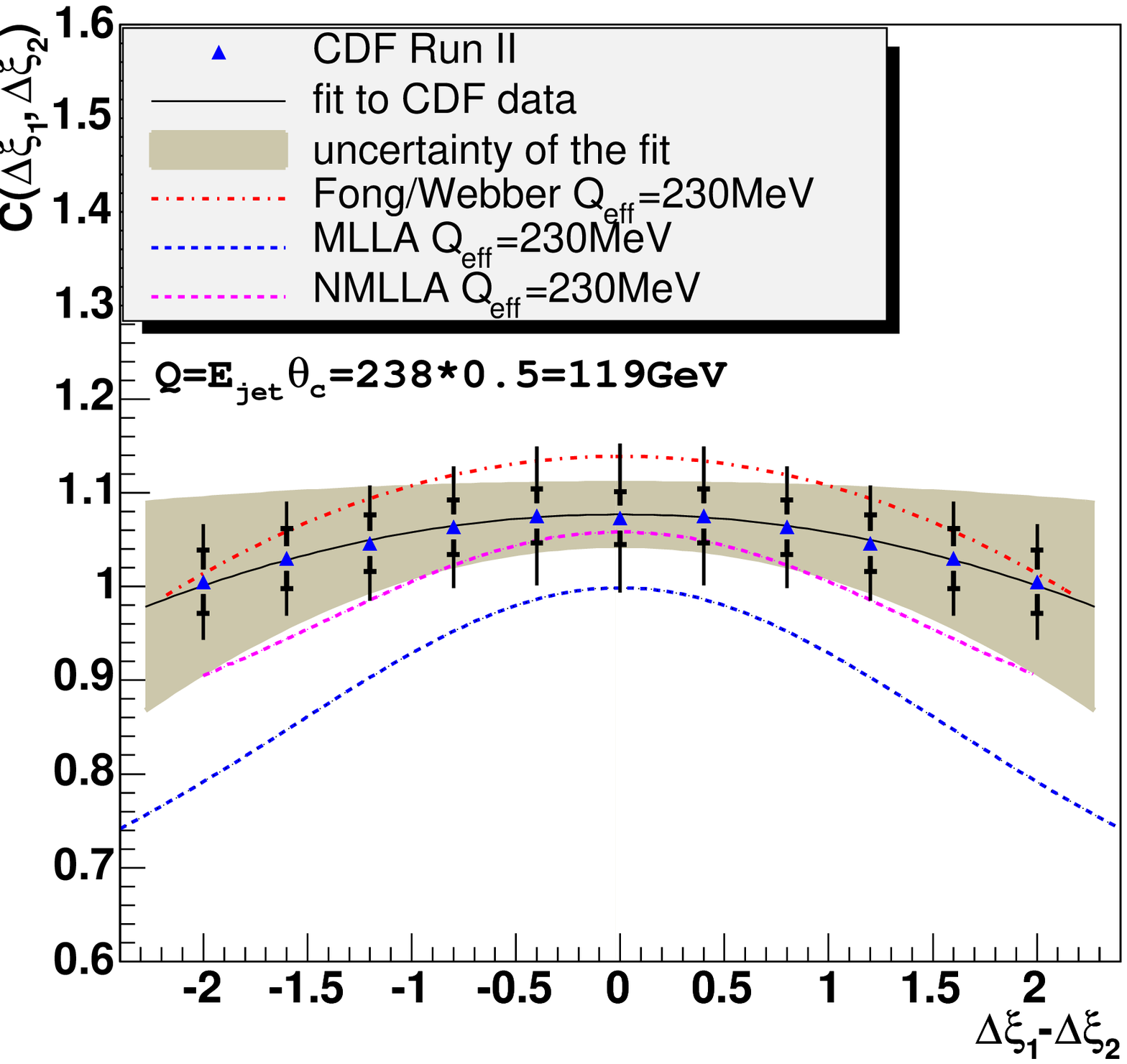}
\caption{\label{fig:cdfcorr} Two-particle correlations in a mixed sample
of gluon and quark jets in $p\bar{p}$ collisions as a function of 
$\Delta\xi_1+\Delta\xi_2=|\ln(x_1x_2)|-\ln(Q/Q_0)$ for $\Delta\xi_1=\Delta\xi_2$ (left)
and $\Delta\xi_1-\Delta\xi_2=|\ln(x_2/x_1)|$ for $\Delta\xi_1=-\Delta\xi_2$ (right).}  
\end{center}
\end{figure}

So far, the MLLA, NMLLA expectations and LPHD for one-particle inclusive 
distributions have been shown to provide a good description of the data.
The study of n-particle correlations is important because, being defined as the 
n-particle cross section normalized by the product of the single inclusive 
distribution (HBP) of each parton, the resulting observable becomes 
independent of the constant ${\cal K}^{ch}$, thus providing a refined test
of QCD dynamics at the parton level.
However, for less inclusive observables like n-particle energy-momentum
correlations and multiplicity correlators, non-trivial hadronization 
effects may appear and spoil the agreement between theory and experiment.
First, the MLLA evolution equations for two-particle correlations, quite similar to
those giving the HBP, were written and solved
iteratively in terms of the logarithmic derivatives of $D^h(\ln(1/x),\ln(k_\perp/Q_0))$ (HBP) \cite{Ramos:2006dx}. 
That is how, the result previously
obtained by Fong and Webber in \cite{Fong:1990nt}, only valid in the vicinity
of the maximum $\ell_{max}$ of the HBP, was extended to all possible values
of $x$. Consequently, as displayed in Fig.\ref{fig:opalcorr}, the normalization of the more accurate solution of the
evolution equations is lower and reproduces some features of the OPAL data at the $Z^0$ peak $Q=91.2$ GeV
of the $e^+e^-$ annihilation, like the flattening of the slopes towards smaller values of $x$ \cite{Ramos:2006dx}. The overall
mismatch between the Fong-Webber and more accurate predictions is of the order ${\cal O}(\alpha_s)$.
Also, the correlation vanishes (${\cal C}^{(2)}\to1$) when one of the partons
becomes very soft, thus describing the hump-backed shape of the one-particle distribution. The reason
for that is dynamical rather than kinematical: radiation of a soft
gluon occurs at large angles which makes the radiation
coherent and thus insensitive to the internal parton structure of the
jet ensemble. Qualitatively, our MLLA expectations agree better
with available OPAL data than the Fong--Webber predictions \cite{Ramos:2006dx}. 
There remains however a significant discrepancy, markedly at very small $x$. In this region
non-perturbative effects are likely to be more pronounced. They may
undermine the applicability to particle correlations of the LPHD 
considerations that were successful in translating parton level predictions 
to hadronic observations in the case of more inclusive single particle energy 
spectra \cite{Dokshitzer:1991wu}. These measurements were redone by the CDF collaboration for $p\bar{p}$ collisions at
center of mass energy $\sqrt{s}=1.96$ TeV for mixed samples of quark and gluon jets \cite{:2008ec}.
For comparison with CDF data, the 2-particle correlator was normalized by the corresponding multiplicity 
correlator of the second rank, which defines the dispersion of the mean average 
multiplicity inside the jet. In this case, the MLLA solution by Fong and Webber \cite{Fong:1990nt}, 
the more accurate MLLA solution \cite{Ramos:2006dx} and the 
NMLLA solution \cite{Arleo:2007wn} were compared with the CDF data. The Fong-Webber predictions turned out to be in good agreement
with CDF data in a range from large to small $x$, also covering the region of the phase space where MLLA 
predictions should normally not be reliable, that is for $x>0.1$, see Fig.\ref{fig:cdfcorr}.
As these figures were taken from \cite{:2008ec}, different notations have been used in this case, 
for instance, $\ell=\xi=\ln(1/x)$, $\Delta\xi=\xi-\xi_{max}$ ($\xi_{max}=\ell_{max}=\frac12\ln(Q/Q_0)$) such 
that $\Delta\xi_1+\Delta\xi_2=\ell_1+\ell_2-\ln(Q/Q_0)$ and $\Delta\xi_1-\Delta\xi_2=\ell_1-\ell_2$.

As observed in Fig.\ref{fig:cdfcorr} (left), 
the data is well described by the three cases in the 
interval $\Delta\xi_1+\Delta\xi_2>-0.5$, that is at very small $x$. However, the Fong and Webber's 
solution also describes the data for $\Delta\xi_1+\Delta\xi_2<-0.5$, that is for larger values of 
$x$ where the MLLA is no longer valid. QCD color coherence for Fig.\ref{fig:cdfcorr} (left,
the peak at $\Delta\xi_1+\Delta\xi_2=-1.5$ is due to numerical uncertainties) should
be observed if the analysis is extended to $\Delta\xi_1+\Delta\xi_2>2.5$.
Moreover, the NMLLA solution \cite{Arleo:2007wn} extends, like for the 
$k_\perp$-spectra, the region of applicability of such predictions for larger values of $x$.
In \cite{:2008ec}, it was concluded that despite the disagreement with the OPAL data in
Fig.\ref{fig:opalcorr}, the LPHD stays successful for the description of less inclusive 
energy-momentum correlations. Therefore, forthcoming data from the LHC becomes necessary
in order to clarify this mismatch. In case the LHC data agrees with CDF, 
the LPHD would stay safe for such observables.
\begin{figure}[h]
\begin{center}
\includegraphics[height=5.5cm,width=7.5cm]{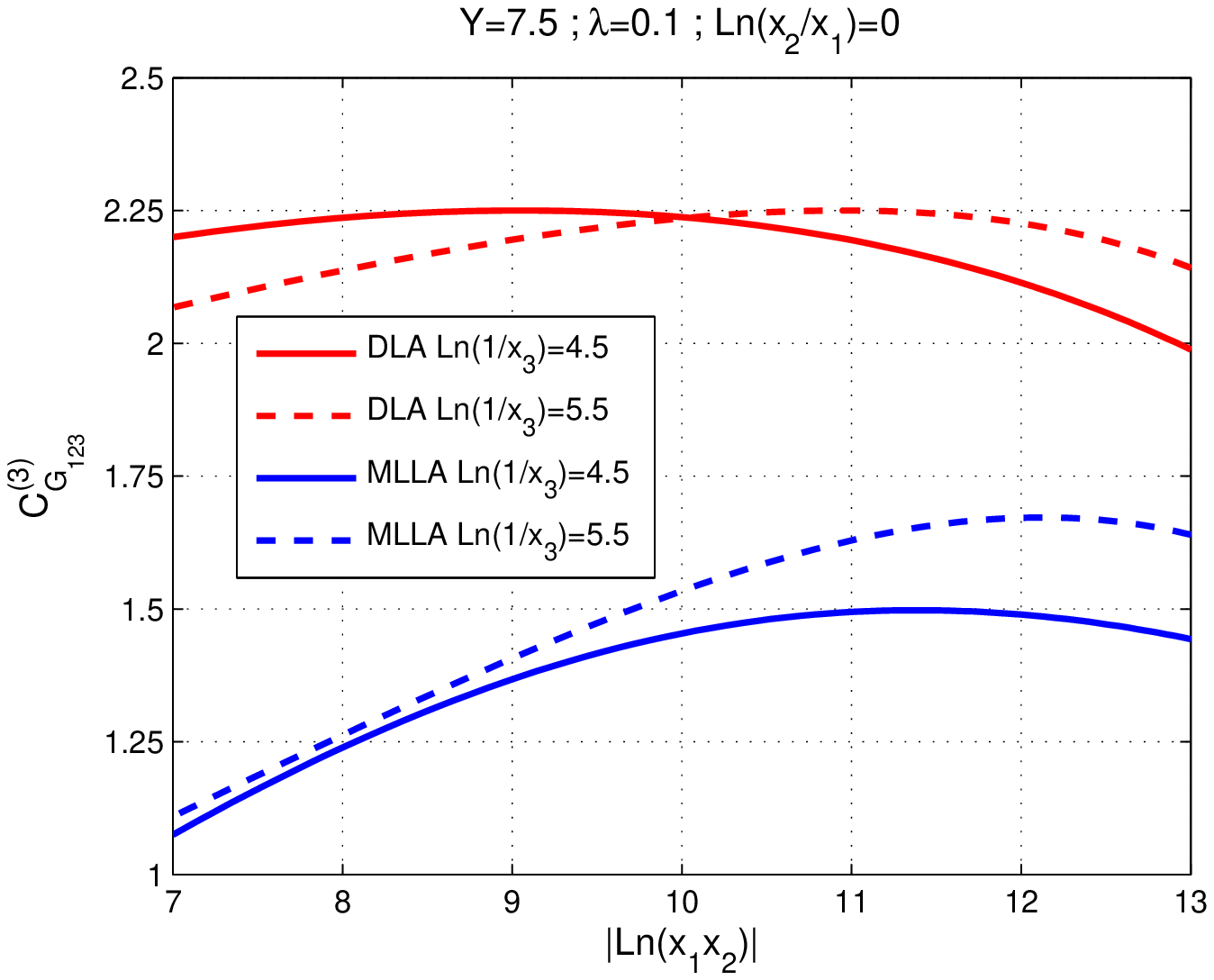}
\includegraphics[height=5.5cm,width=7.5cm]{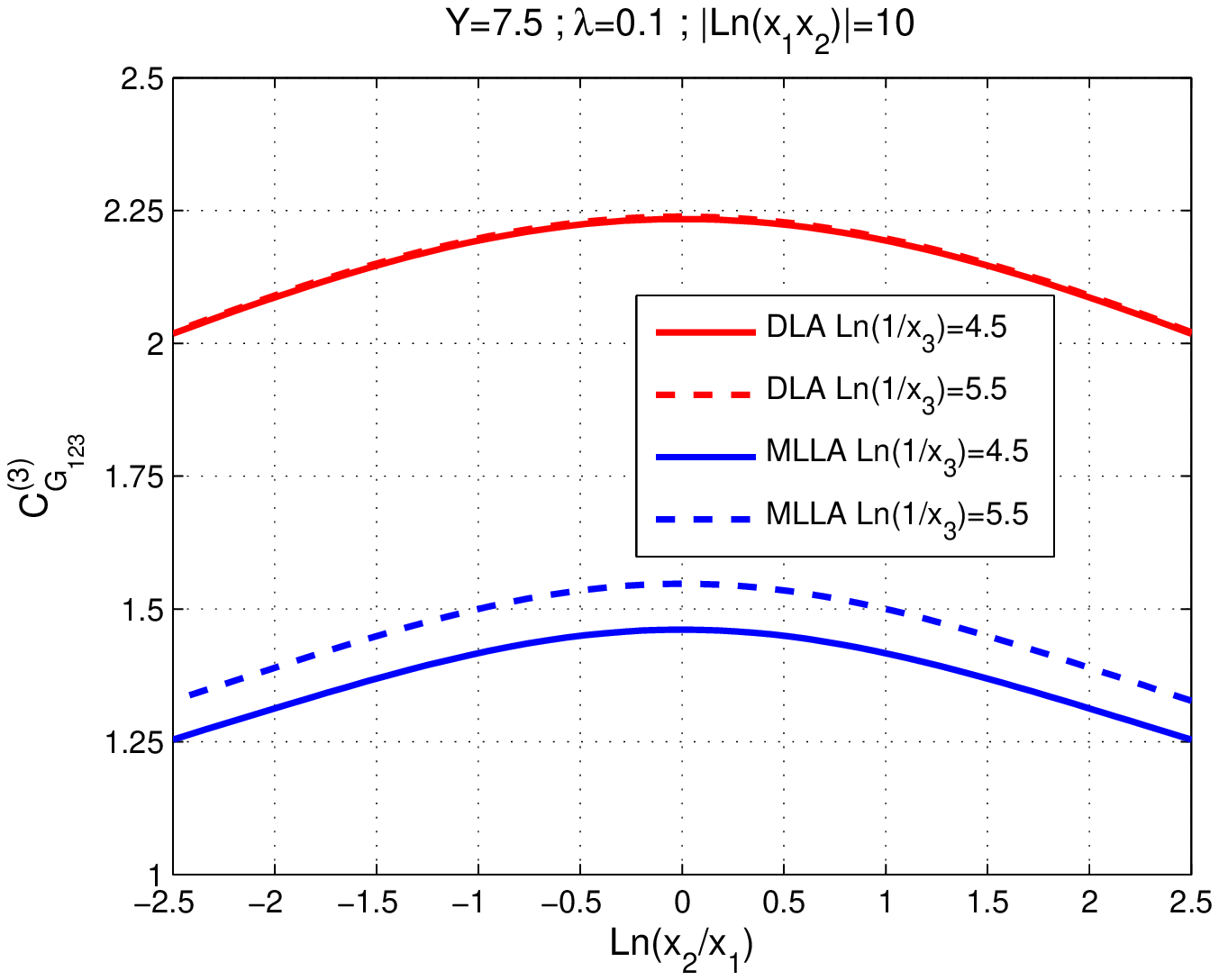}
\caption{\label{fig:3partcorr} Gluon jet 3-particle correlator as a function of 
$|\ln(x_1x_2)|$ for $x_1=x_2$ 
and $\ln(1/x_3)$ and as a function of
$\ln(x_2/x_1)$ for fixed $|\ln(x_1x_2)|$ 
and $\ln(1/x_3)$ (right).}  
\end{center}
\end{figure}

Finally, we perform theoretical predictions for three-particle correlations 
for the LHC in the limiting spectrum approximation ($Q_0\approx\Lambda_{QCD}$) \cite{Ramos:2011tw}.
The correlators are functions of the total hardness $Q$ of the jet and the three $x_i$ ($i=1,2,3$)
energy fractions: ${\cal C}^{(3)}_{G_{123}}(\ell_1,\ell_2,\ell_3,Y)$ 
and ${\cal C}^{(3)}_{Q_{123}}(\ell_1,\ell_2,\ell_3,Y)$.
In Fig.\ref{fig:3partcorr}, the DLA and 
MLLA three-particle correlators inside a gluon jet are displayed, 
as a function of the difference $(\ell_1-\ell_2)=\ln(x_2/x_1)$
for two fixed values of $\ell_3=\ln(1/x_3)=4.5,\,5.5$, fixed sum $(\ell_1+\ell_2)=|\ln(x_1x_2)|=10$ 
and, finally, fixed $Y=7.5$ (virtuality $Q=450$ GeV and $\Lambda_{QCD}=230$ MeV), which is realistic for LHC 
phenomenology. The representative values $\ell_3=\ln(1/x_3)=4.5,\,5.5$ ($x_3=0.011,\,x_3=0.004$) 
have been chosen according to the range of the energy fraction $x_i\ll0.1$, 
where the MLLA scheme can only be applied. From Fig.\ref{fig:3partcorr}, the gluon
correlator is observed to be the strongest when particles have the same energy and 
to decrease when one parton is harder than the others, which as for the one and two-particle
distributions, follows as a consequence of coherence. 
Moreover, the observable increases for softer partons with 
$x_3$ decreasing, which is for partons less sensitive to the energy balance. This
observable becomes useful so as to extend the domain of applicability of the LPHD such that it 
can be measured together with two-particle correlations at the LHC. Same conclusions and trends 
hold for the quark correlator but the normalization is higher, showing stronger 
correlations inside a quark jet \cite{Ramos:2011tw}.
Finally, after jets are properly reconstructed, the measurement of the one-particle
spectrum, two and three-particles correlations becomes extremely important at the LHC in order
to further test the LPHD hypothesis.

\end{document}